\documentclass[conference]{IEEEtran}
\IEEEoverridecommandlockouts
\ifCLASSINFOpdf
  \usepackage[pdftex]{graphicx}
\else
\fi
\usepackage[cmex10]{amsmath}
\usepackage{array}

\usepackage[tight,footnotesize]{subfigure}
\begin{document}
%
\title{Conceptual Design of a New Large Superconducting Toroid for IAXO,\\ the New International AXion Observatory
}



\author{I. Shilon, A. Dudarev, H. Silva and H. H. J. ten Kate

\thanks{Manuscript received October 9, 2012.}
\thanks{I. Shilon, A. Dudarev, H. Silva and H. H. J. ten Kate are with the European Organization for Nuclear Research (CERN), CH-1211, Gen\`eve 23, Switzerland (phone: 0041-22-767-0963; e-mail: idan.shilon@cern.ch).}
\thanks{I. Shilon is also with the Laboratorio de F\'{\i}sica Nuclear y Astroparticulas, Universidad de Zaragoza, Spain.}}


%



\maketitle

\begin{abstract}

The International AXion Observatory (IAXO) will incorporate a new generation detector for axions, a hypothetical particle, which was postulated to solve one of the puzzles arising in the standard model of particle physics, namely the strong CP (Charge conjugation and Parity) problem. The new IAXO experiment is aiming at achieving a sensitivity to the coupling between axions and photons of one order of magnitude beyond the limits of the current state-of-the-art detector, represented by the CERN Axion Solar Telescope (CAST). The IAXO detector relies on a high-magnetic field distributed over a very large volume to convert solar axions into x-ray photons. Utilizing the designs of the ATLAS barrel and end-cap toroids, a large superconducting toroidal magnet is currently being designed at CERN to provide the required magnetic field. The new toroid will be built up from eight, one meter wide and 20~m long, racetrack coils. The toroid is sized about 4~m in diameter and 22~m in length. It is designed to realize a peak magnetic field of 5.4~T with a stored energy of 500~MJ.
The magnetic field optimization process to arrive at maximum detector yield is described. In addition, force and stress calculations are performed to select materials and determine their structure and sizing. Conductor dimensionality, quench protection and the cryogenic design are dealt with as well.
\end{abstract}


%
\IEEEpeerreviewmaketitle

\section{Introduction}

The strong CP (Charge conjugation and Parity) problem is the puzzling question of why quantum chromodynamics (QCD) does not seem to break the CP symmetry. An attractive solution to this puzzle invokes a new $U(1)$ symmetry, the so-called Peccei-Quinn (PQ) symmetry \cite{Peccei:1977hh, Peccei:1977ur}. Associated with the spontaneous breaking of the PQ symmetry is a light neutral pseudoscalar particle, the axion, which is closely related to the neutral pion \cite{Weinberg:1977ma, Wilczek:1977pj}. Axions are one of the most interesting non-baryonic candidates for dark matter in the universe. They may also exist as primordial cosmic relics copiously produced in the early Universe. For these reasons, axions have received much attention.

\renewcommand{\arraystretch}{1.0825}
\begin{table}[htdp]
\caption{main design parameters of IAXO's toroidal magnet}
\begin{center}
\begin{tabular*}{0.4\textwidth}{@{\extracolsep{\fill}} p{5.35 cm}  c } \hline
\textit{Property} & \textit{Value}\\ 
\hline
\textbf{Size:} \hfill Inner radius, $R_{in}$ (m) & 1.05  \\
\hfill Outer radius, $R_{out}$ (m)& 2.05 \\
\hfill Total axial length (m) & 22\\
\textbf{Mass:} \hfill Conductor (tons) & 65\\
\hfill Cold Mass (tons) & 130\\
\hfill Cryostat (tons) & 35\\
\hfill Total assembly (tons) & $\sim$250\\
\textbf{Coils:} \hfill Turns/coil & 180\\
\hfill Nominal current, $I_{op}$ (kA) & 12\\
\hfill Stored energy, $E$ (MJ) & 500\\
\hfill Peak magnetic field, $B_p$ (T) & 5.4\\
\hfill Average field in the bores (T) & 2.5 \\
\textbf{Conductor:} \hfill Overall size (mm$^2$) & 35 $\times$ 8\\
\hfill Number of strands & 40\\
\hfill Strand diameter (mm) & 1.3\\
\hfill Operational margin & 40\%\\
\hfill Temperature margin @ 5 T (K) & 1.9\\
\textbf{Heat Load:} \hfill  at 4.5 K (W) & $\sim$150\\
\hfill at 60-80 K (kW) & $\sim$1.6\\
\hline \end{tabular*} 
\end{center}
\label{table1}
\end{table}

\begin{figure}[!h]
\begin{center}
    \includegraphics[scale=0.54] {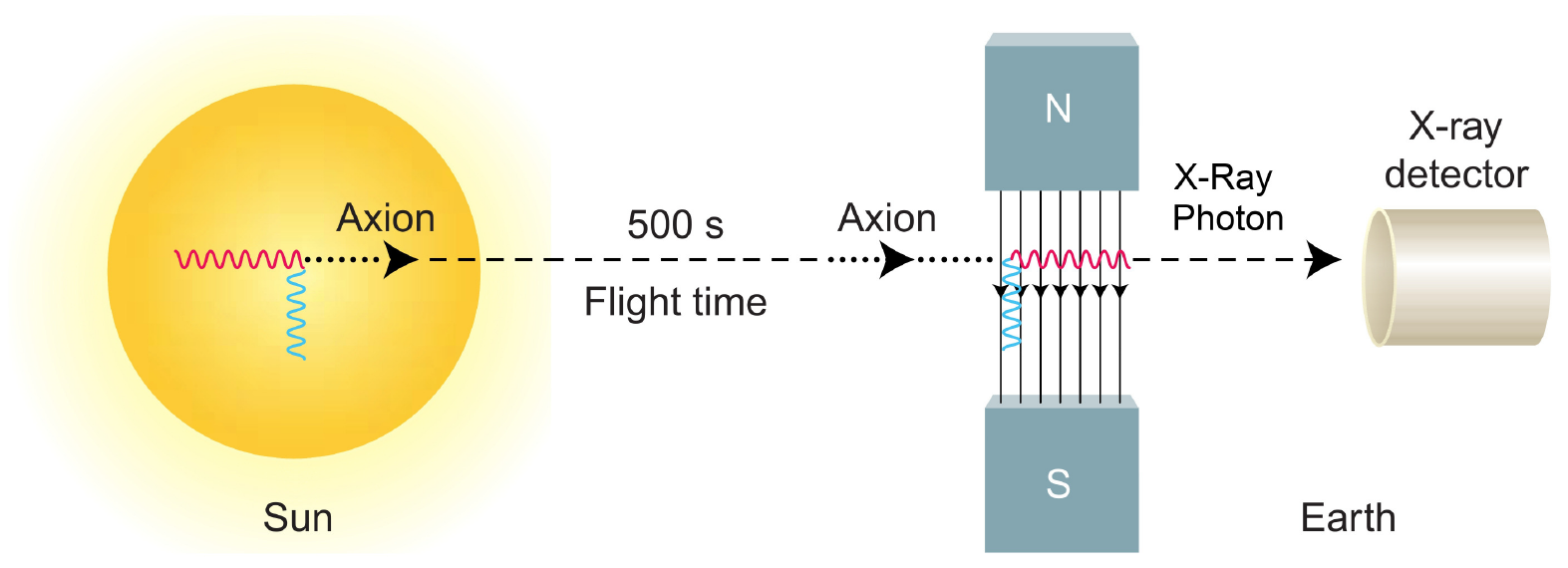}
    \caption{Schematic view of the solar axions detection concept: solar axions are predicted to convert to detectable x-ray photons when interacting with a magnetic field. Source: Science magazine Vol. 308 (2005).}
    \label{fig:1}
\end{center}
\end{figure}

Axions interact very weakly with ordinary matter, thus being practically "invisible" particles. However, axions are predicted to convert to and from photons in the presence of a high-magnetic field. When an axion travels through a magnetic field region, it interacts with this magnetic field and may convert to a detectable photon (see Fig. \ref{fig:1}). This property is used to detect axions in terrestrial experiments. It is the aim of this work to define a new superconducting toroidal detector magnet that will be specifically designed for solar axions detection and will constitute the backbone of IAXO, the International AXion Observatory. The IAXO project entails an immense upgrade of axion detection experiments, compared to the current state-of-the-art which is represented by the CAST experiment at CERN and using a 9~T, 9~m long, LHC dipole prototype magnet with twin $15$~cm$^2$ bores. IAXO features a dramatic expansion of the present limits on axions search.

\section{Design Concept}

\subsection{Figure of Merit and Lay-Out Optimization}

The first step towards a conceptual design of IAXO's magnet is to optimize its toroidal geometry in order to meet the main design criterion, namely, achieving a sensitivity to the coupling between axions and photons of one order of magnitude beyond the limit of the current experiment (i.e. CAST). The magnet's figure of merit (MFOM) is given by $f_M = L^2B^2A$ \cite{IAXO}, where $L$ is the magnet's length, $B$ the magnetic field and $A$ the aperture covered by the x-ray optics. Currently, CAST's MFOM is 21~T$^2$m$^4$. Translating the design criterion to the $f_M$ language, the IAXO magnet should achieve an MFOM of 300, relative to CAST. One should notice that the complete figure of merit of the experiment includes the tracking, detectors and optics parameters as well and is given in \cite{IAXO}.

\begin{figure}[!h]
\begin{center}
	\subfigure{\includegraphics[scale=0.088]{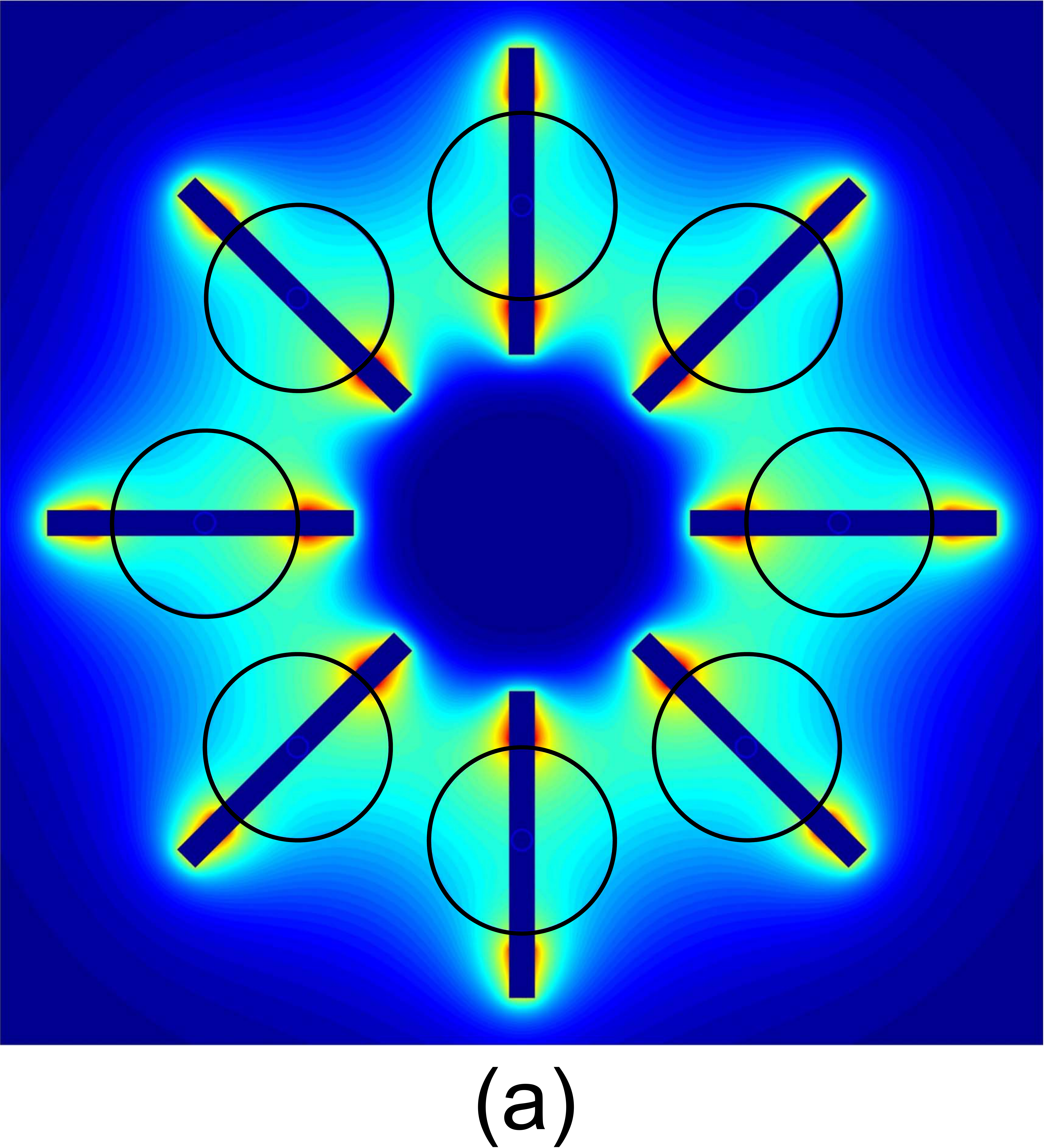}}
	\subfigure{\includegraphics[scale=0.088]{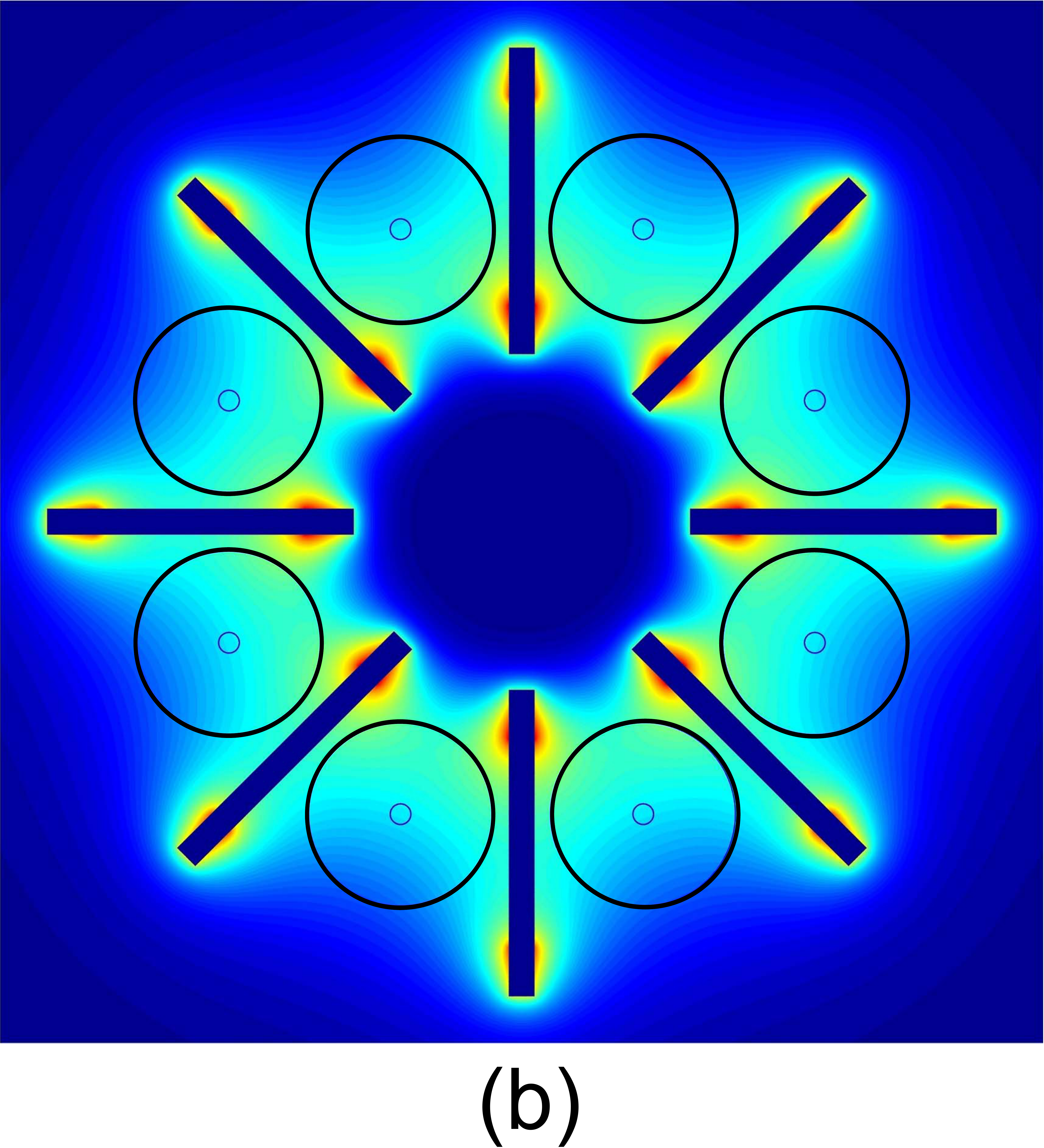}}
	\caption{Illustration of the two principle angular alignment options considered for the telescopes with respect to the coils. The rectangles represent the toroid's coil and the circles represent the telescopes' bores. (a) "Field dominated" alignment: telescopes behind the coils. (b) "Area dominated" alignment: telescopes between the coils.} 
	\label{alignment} 
\end{center}
\end{figure}

When the magnet's straight section length $L$ is set to 20 m, the MFOM is determined by the integral $\int B^2(x,y)dxdy$. The integration is performed solely across the \textit{open} area covered by the x-ray optics. Hence, to perform the integration, the telescopes' positioning must be determined. Upon minimizing the radial positioning, two principal options for the angular alignment are considered: one is to align each of the telescopes between each pair of racetrack coils, whereas the other is to place the telescopes behind the racetrack coils. Fig. \ref{alignment} provides a general illustration of the two alignment options for an eight coils toroid. In practice, the two options represent two different approaches: the first, referred to as the "area dominated" option, takes advantage of the entire large aperture of each of the telescopes and the second "field dominated" option assumes that placing the coils behind the optics, and by that including areas with higher magnetic field in the integration, will increase $f_M$. 

Once the telescopes' position is fixed, one can perform the integration over a disc with radius $R_{det}$ centered at $(R_{cen}, \theta_{cen})$. Then, the magnetic field is determined by the magnet's geometrical and electromagnetic parameters. For each lay-out configuration, the magnetic field is calculated using a 3D FEA model and the integration is then performed on the mid-plane. The model features an arc at the bended sections of each racetrack with a radius $R_{arc} = (R_{out} - R_{in})/2$, where $R_{out}$ and $R_{in}$ are the outer and inner radii of the racetrack coils, respectively. The model also assumes the use of an Al stabilized Rutherford NbTi cable in the coil windings. The winding dimensions are determined from the conductor dimensions assuming a few different winding configurations.

The optimization study shows that IAXO's MFOM is effected considerably by the fraction of the telescopes' aperture exposed to x-rays. Even when considering the "field dominated" alignment, it is preferable to use thinner coils, thus increasing the open aperture that is picked up by the telescopes. Moreover, the "area dominated" option gives a $15 \%$ larger MFOM, compared to the "field dominated" option. 

\begin{figure}[!b]
\begin{center}
    \includegraphics[scale=0.102] {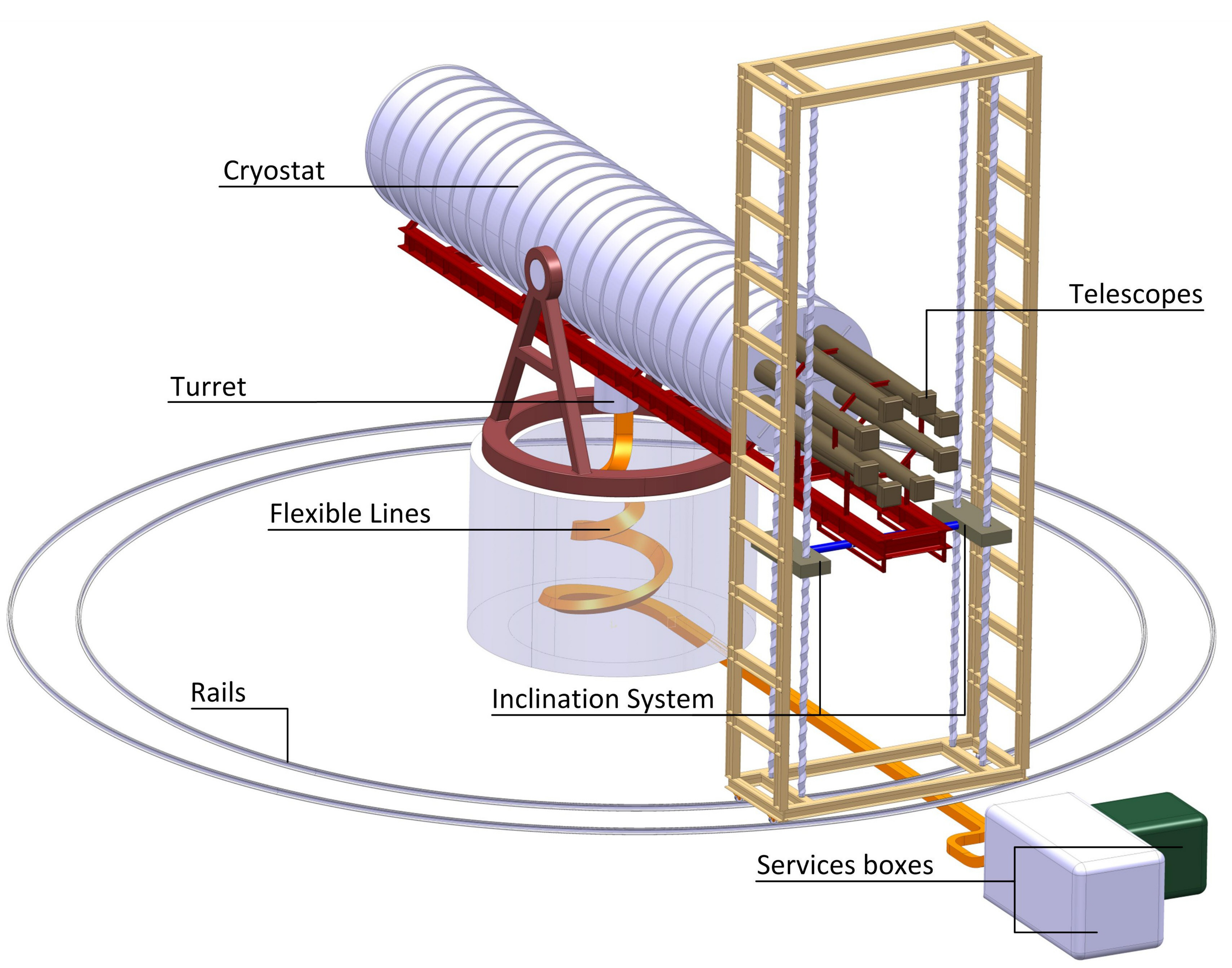}
    \caption{Schematic view of the IAXO system. Shown are the cryostat, eight telescopes, turret leading the flexible services lines into the magnet, cryogenics and powering services boxes, inclination system and the rails for horizontal movement.}
    \label{fig:2}
\end{center}
\end{figure}

The result of the geometrical optimization study leads to a design concept, as presented in Fig. \ref{fig:2}, addressing all the experimental requirements of the magnet while relying on known and mostly well proven engineering solutions. This leads to a magnet design which is technically feasible to manufacture. Furthermore, the design concept essentially supports the decoupling of the magnet system from the optical detection systems. This allows for open bores, which are centered and aligned in between the racetrack coils, in accordance to the geometrical study. The eight warm bores will simplify the use of physics experimental instrumentation and the regular maintenance of the system. The diameter of the bores matches the diameter of the telescopes in use.

The toroidal magnet comprises the coil and its casing, an inner cylindrical support for the magnetic forces, a thermal shield, a vacuum vessel and a movement system (see Figs. \ref{fig:2} to \ref{cs} and Table I). Its mass is $\sim$250~tons with a stored energy of $\sim$500~MJ. Throughout the structural design study, the design goals are: a maximum deflection of 5 mm, a general stress limit of 50~MPa and a buckling factor of 5. The magnetic and structural designs are carried using the ANSYS$^{\textregistered}$ Workbench environment. The Maxwell 15.0 code is used to calculate 3D magnetic fields and Lorentz forces. The magnetic force load is then transferred to Workbench 14.0 to calculate stress and deformation. The eight bores in between the racetrack coils are facing eight x-ray telescopes with a diameter of 600 mm, which implies a focal length of 5 m. The choice for an eight coils toroid with the given dimensions and eight 600 mm telescopes and bores is also determined by cost optimization within the anticipated budget.  

We end this section by mentioning that numerous magnet designs (e.g. accelerator magnets, solenoids and dipole structures) were considered during the optimization study until the conclusion that the optimal axion helioscope geometry has toroidal symmetry was made \cite{IAXO}. In addition, less conventional toroidal designs were also examined. For example, ideas for racetrack windings with bended ends were suggested to reduce the area loss when implementing the field dominated alignment. For the same reason toroids with slightly tilted coils were also discussed. In general, these designs posses technical complications while offering a low potential to significantly enlarge the MFOM and hence deviate from the philosophy behind our magnet concept. In addition, toroidal geometries with different number of coils were also studied. The MFOM scales linearly with the number of coils (when keeping the telescopes' cross-section constant), which points out that, as mentioned, the choice for an eight toroid is cost driven in essence. 

\subsection{Conductor}

\begin{figure}
\begin{center}
    \includegraphics[scale=0.28] {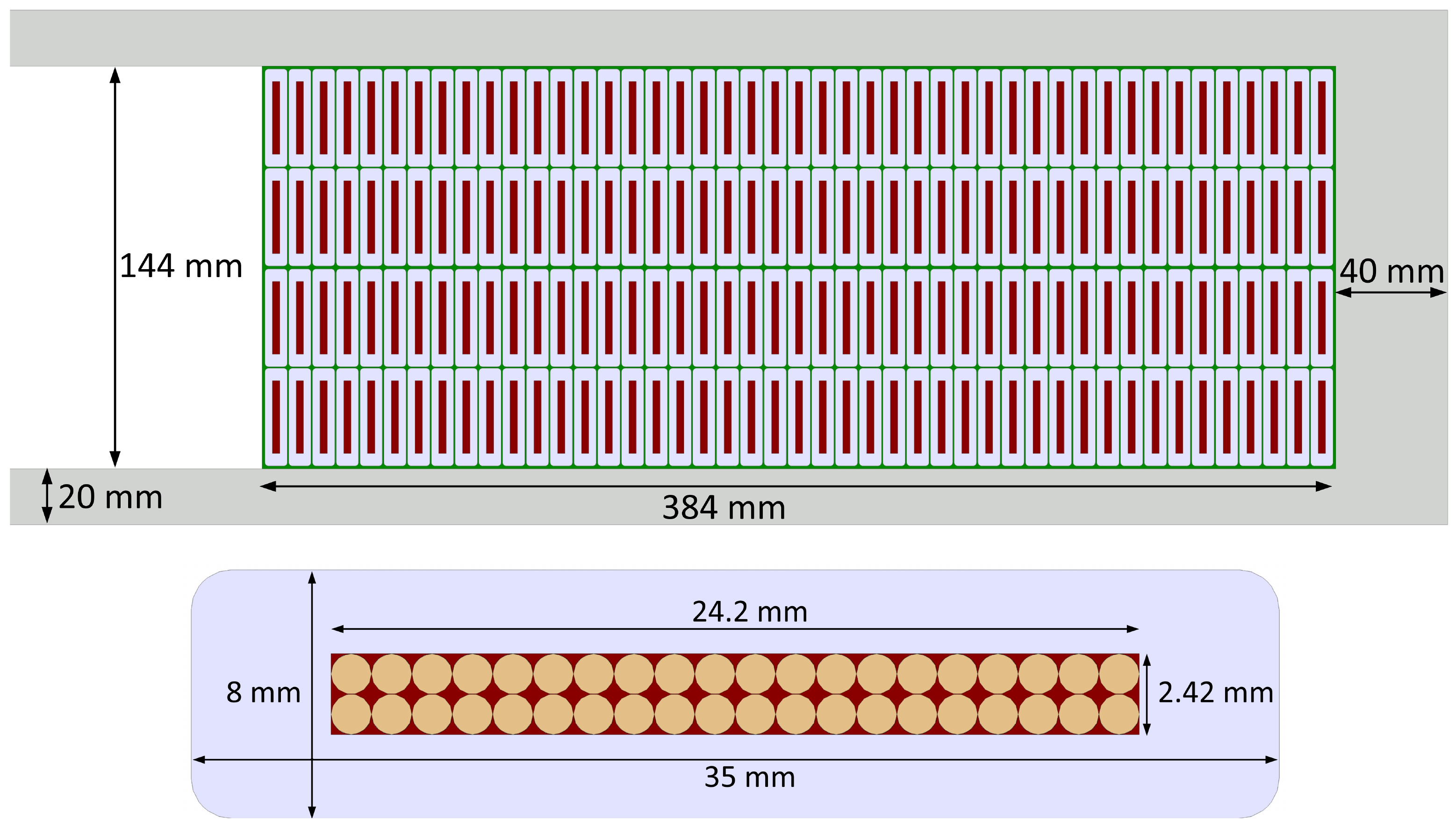}
    \caption{Cross section of the conceptual design of the two double pancake winding pack and the coil casing (top) and the conductor with a 40 strands NbTi Rutherford cable embedded in a high purity Al stabilizer (bottom).}
    \label{coil}
\end{center}
\end{figure}

The conductor design is based on a large Rutherford type NbTi cable (with 40 strands of 1.3 mm diameter and Cu/SC ratio of 1.1) embedded in a high-purity aluminum stabilizer, following the techniques used in the ATLAS and CMS experiments at CERN \cite{ATLAS}-\cite{CMS} (see Fig. \ref{coil}). The use of the Rutherford cable as the superconducting element provides high current density while maintaining high performance redundancy in the large number of strands. The Al stabilizer serves both quench protection and stability for the superconductor.

\subsubsection{Peak Magnetic Field and Forces}

The peak magnetic field in the windings, which determines the operation point of the conductor and the temperature margin, is calculated at 5.4~T for a current of 12~kA per turn. In order to minimize the forces acting on the bended sections, the racetrack coils are bent to a symmetric arc shape, with $R_{arc} = 0.5$~m. The net force acting on each racetrack coil is 18.8~MN, directed radially inwards.

\subsubsection{Stability Analysis}

The IAXO magnet requires maintaining the highest possible magnetic field in order to maximize the MFOM. However, acceptable operational and temperature margins are still required to ensure proper operation. For a two double pancake configuration with 180 turns and engineering current density $J_{eng} = 40$~A/mm$^2$, the peak magnetic field is $B_p = 5.4$~T, whereas the magnetic field on the inner edge of the coils at the toroid's mid-plane (where the $f_M$ integration is performed) is $B_{mid} = 5$~T. Then, we find that $J_c = 65$~A/mm$^2$, where the corresponding peak magnetic field is 8.8~T. Hence, IAXO's magnet is expected to work at about 60\% on the load line, which sets the operational margin to 40\%.

The temperature margin calculation is carried out with $T_{op} = 4.5$~K and for $B_p = 5.4$~T (which corresponds to a configuration with $B_{mid} = 5$~T). The number of turns in the coil is varied, aiming for approximately a 2~K temperature margin. A coil with two double pancakes and 45 turns per pancake satisfies this requirement with a temperature margin of 1.9~K, while yielding an MFOM of about 300, relative to CAST, which meets the principal design criterion.


\subsection{Quench Protection}

The adiabatic temperature rise due to a uniformly distributed quench is about 100~K. Therefore, the quench protection is based on an active protection system with multiple heaters along each of the eight racetracks, to enforce a homogenous cold mass temperature after a quench.

\subsection{Cold-Mass}

The cold mass' operating temperature is 4.5~K and its weight is approximately 130~tons. It consists of eight coils, with two double pancakes per coil, which form the toroid geometry and a central cylinder designed to support the magnetic force load. The coils are embedded inside Al5083 casings, which are attached to the support cylinder at their inner edge. The casings are designed to prevent coil deflection due to the magnetic forces.

A coil, shown in Fig. \ref{coil}, comprises two double pancake windings, separated by a 1 mm layer of insulation. The coils will be impregnated for final bonding and pre-stressed within their individual casing to minimize shear stress and prevent cracks and gaps appearing due to magnetic forces. 

\begin{figure}[b!]
\begin{center}
    \includegraphics[scale=0.375] {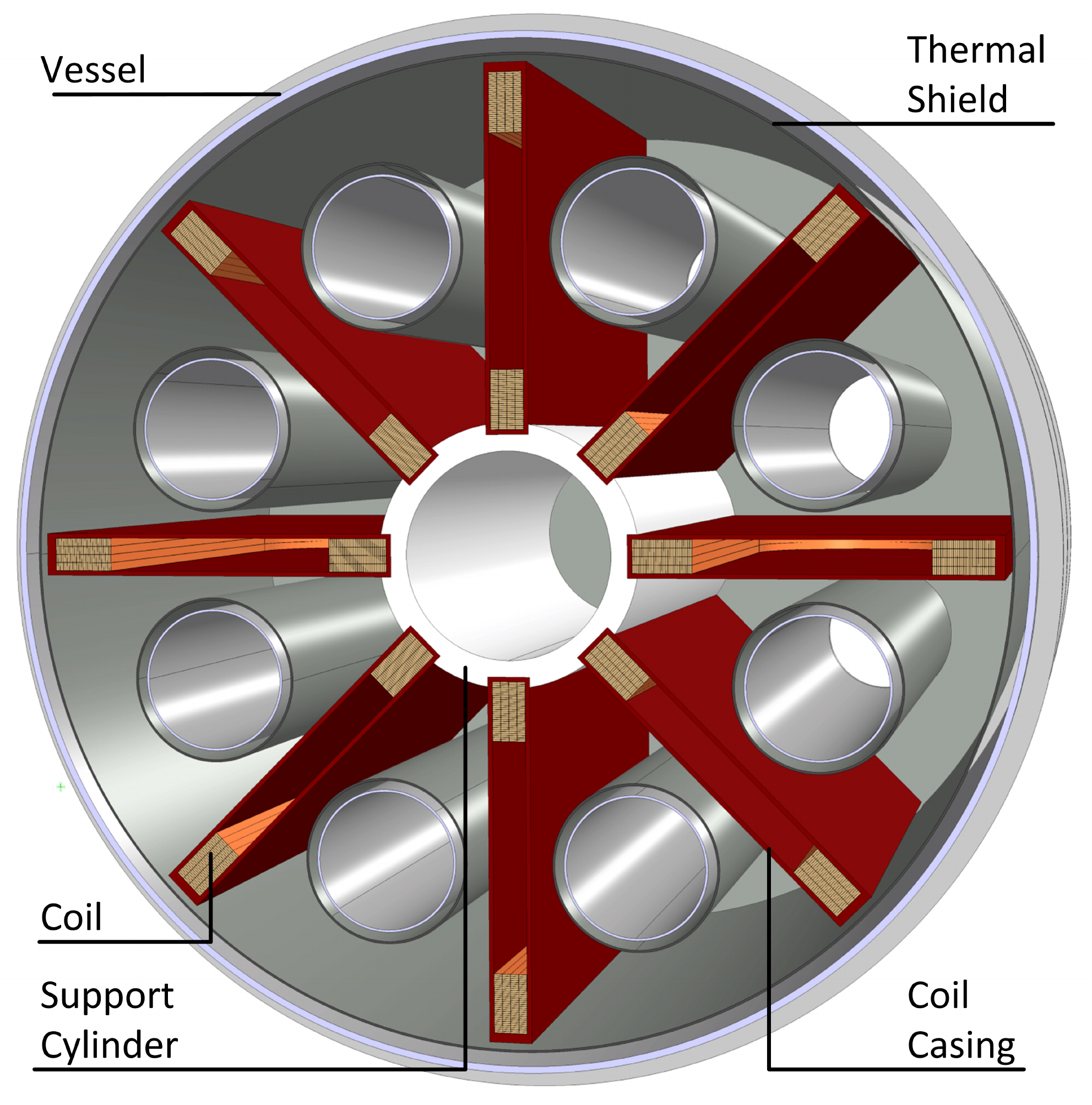}
    \caption{Mid-plane cut of the magnet vessel with depth view, showing the cold mass, surrounded by a thermal shield, and the cryostat. The open bores will allow for easy regular maintenance of the magnet system while simplifying the use of experimental instrumentation.}
    \label{cs}
\end{center}
\end{figure}

\subsection{Cryostat and Movement System}

The design of the cryostat is based on a large cylinder and two thick end plates with eight cylindrical open bores placed in between them, see Fig. \ref{cs}. The vessel is optimized to sustain the atmospheric pressure difference and the gravitational load, while being as light and thin as possible. Using two end flanges at the vessel's rims, as well as periodic reinforcement ribs at one meter intervals along the cryostat's length, the structural design goals are met for a 10~mm thick Al5083 vessel with two 70~mm thick end plates. The 10~mm wall thickness of the eight cylindrical bores have also been minimized in order for the bores to be placed as close as possible to the racetracks' inner radius, thereby increasing the MFOM. 


Searching for solar axions, the IAXO detectors need to track the sun for the longest possible period in order to increase the data-taking efficiency. Thus, the magnet needs to be able to rotate both horizontally and vertically by the largest possible angles. For vertical inclination a $\pm$ 25$^\circ$ movement is required, while the horizontal rotation should be stretched to a full 360$^\circ$ rotation before the magnet revolves back at a faster speed to its starting position. 

The 250 tons magnet system will be supported at its center of mass point (see Fig. \ref{fig:2}), thus minimizing the torques acting on the support structure and allowing for simple rotation and inclination mechanisms. Accordingly, a set of drives can be used to rotate the magnet, while a simple inclination structure can be placed behind the detectors (the farthest part of the system from its center of mass point), minimizing the force required to incline the magnet. The magnet services can be connected via a turret aligned with the rotation axis, thus simplifying the flexible cables and transfer lines arrangement. A 3D spiral shaped flexible chain, with radius and length altering as the magnet moves, is guiding the services lines and cables to the stationary connection point.

\subsection{Cryogenics}

The coil windings are cooled by liquid helium at a temperature of 4.5 K via conduction. The cooling scheme utilizes a forced flow helium cooling, where the coolant flows in a pipe system which will be attached to the coil casings, thus allowing for conduction cooling, in a manner similar to the ATLAS toroids \cite{ATLAS, ATLAS2}. The decision for using this concept is following the same philosophy of our concept: a known technology with a low-cost proven solution and most reliable.

The heat load on the magnet by radiation and conduction is $\sim$150~W at 4.5~K. In addition comes the thermal shield heat load of $\sim$1.6~kW. An acceptable thermal design goal is to limit temperature rise in the coils to 0.1~K above the coolant temperature under the given heat loads. To warrant this, a mass flow rate of $\sim$ 200~g/s is required. 

A challenge is to maintain a stable forced flow against gravity because the magnet has to rotate at both positive and negative vertical angles. A possible solution is to use a dual cooling system, driven by a cold He-pump and a proper set of valves. When the magnet passes through its horizontal while tracking the sun, the helium flow is switched from one to the other circuit, thereby guaranteeing continuous helium flow along the coil casings.

\section{Conclusion}

The design of the new IAXO toroidal superconducting magnet satisfies the design criterion of increasing the solar axion searches sensitivity by at least 1 order of magnitude. The design is relying on known engineering solutions and manufacturing techniques. The design features maximum flexibility for the experimentalists, thus allowing to maximize and extend the detection potential of not only the search of axions, but also of more exotic particles, commonly known as Axion Like Particles (ALPs). Essentially, the magnet system and optical detectors are separated, allowing for a parallel effort of development, construction and initial operation. This, thereby, also minimizes cost and risks. In the coming years the design will be further detailed and project funding will be proposed.



\section*{Acknowledgment}

The authors would like to thank D. Mladenov of CERN for thorough and useful discussions on FE analysis. We acknowledge L. Walckiers and S. Russenschuck of CERN for directing the project to its current course at its early stages. We thank I. Ortega for his contribution to the lay-out optimization study. I. Shilon acknowledges the support from the Spanish Ministry of Economy and Competitiveness (MINECO) under contract FPA2011-24058. H. Silva acknowledges the support from the Portuguese Foundation for Science and Technology (FCT) under contract SFRH/BEST/51761/2011. Lastly, we thank our IAXO collaborators for motivating this work.



%

\end{document}